\documentclass[5p,twocolumn,times,number]{elsarticle}

\usepackage{graphicx}
\usepackage{amsmath}   

\newcommand{\tabref}[1]{Tab.~(\ref{#1})}

\begin{document}

\begin{frontmatter}

\title{Development of the superconducting detectors and read-out for the X-IFU instrument on board of the X-ray observatory Athena}

\author[add1]{L.~Gottardi\corref{cor}}
\ead{l.gottardi@sron.nl}
\author[add1]{H.~Akamatsu}
\author[add1]{M.~Bruijn}
\author[add1]{R.~den Hartog}
\author[add1]{J.-W.~den Herder}
\author[add1]{B.~Jackson}
\author[add2]{M.~Kiviranta}
\author[add1]{J.~van der Kuur}
\author[add1]{H.~van Weers}

\cortext[cor]{Corresponding author}

\address[add1]{SRON Netherlands Institute for Space Research, Utrecht, The Netherlands}
\address[add2]{VTT, Espoo, Espoo, Finland}

\begin{abstract}
The Advanced Telescope for High-Energy Astrophysics (Athena) has been selected by ESA as its second large-class mission. The future European X-ray observatory will study the hot and energetic Universe with its launch foreseen in 2028. 
Microcalorimeters based on superconducting  Transition-edge sensor
(TES)  are the chosen technology for the detectors array of the X-ray
Integral Field Unit (X-IFU) on board of Athena. The X-IFU is a 2-D
imaging integral-field spectrometer operating in the soft X-ray band
($0.3 - 12\, \mathrm{keV}$). The detector consists of an array of 3840 TESs coupled to X-ray absorbers and read out in the MHz bandwidth using Frequency Domain Multiplexing (FDM) based on Superconducting QUantum Interference Devices (SQUIDs). The proposed design calls for devices with a high filling-factor, high quantum efficiency, relatively high count-rate capability and an energy resolution of 2.5 eV at 5.9 keV.
The paper will review the basic principle and the physics of the TES-based microcalorimeters and present the state-of-the art of the FDM read-out.

\end{abstract}


\end{frontmatter}


\section{Introduction}
The Advanced Telescope for High-ENergy Astrophysics ({\it Athena})
will be an X-ray telescope designed to address the Cosmic Vision
science theme 'The Hot and Energetic Universe' and its launch is
foreseen in 2028.  It will answer important science questions like\cite{Barcons:2015}:
\begin{itemize}  
 \item How does ordinary matter assemble into the large-scale
   structures we see today?
 \item How do black holes grow and shape the Universe?
\end{itemize}

Thanks to  its revolutionary optics  technology and to the  most advanced
X-ray  instrumentation, Athena  will  have superior  wide field  X-ray
imaging,  timing  and imaging  spectroscopy  capabilities, far  beyond
those  of any  existing observatory.   The X-ray  Integral  Field Unit
(X-IFU) is one of  the two instrument on board of Athena.  It is a 2-D
imaging integral-field  spectrometer operating in the  soft X-ray band
($0.2-12\, \mathrm{keV}$). It will provide breakthrough capabilities for mapping in
3D  the hot  cosmic gas  to study  for example  the process  of matter
assembly in clusters and for  detecting weak lines to characterise the metals  in clusters of galaxies or  the missing baryons in
the Warm-Hot Intergalactic Medium \cite{Ravera:2014}.
The X-IFU proposed design calls for devices with an high filling-factor, high quantum efficiency,
relatively high count-rate capability and energy resolution better than  3 eV at 5.9 keV. 
The major requirements for the X-IFU instrument are listed in \tabref{xifutable}.
\begin{table}[htbp]
  \begin{center} \small
 
    \begin{tabular}{l l}  
\hline
\vspace{0.06 cm} {\bf Parameters} & {\bf Requirements}  \\
\hline
\vspace{0.1 cm} Energy range & $0.2-12\,\mathrm{keV}$\\
\vspace{0.1 cm} Energy resolution: $E < 7\, \mathrm{keV}$ & $2.5\, \mathrm{eV\
}$ \\
\vspace{0.1 cm} Pixel size &  $250 \times 250\, \mu \mathrm{m}^2$\\
\vspace{0.1 cm} Field of view & $5^\prime$ (diameter)\\
\vspace{0.1 cm} Quantum efficiency @ $6\, \mathrm{keV}$ & $>90\%$\\
\vspace{0.1 cm} Count rate - faint source & 1 mCrab ($>80\%$ high-resolution)\\
\vspace{0.1 cm} Count rate - bright source & 1 Crab ($>30\%$ low-resolution)\\
\vspace{0.1 cm} non X-ray background&  ($<5 \times10^{-3}\, \mathrm{cts/cm}^2/\mathrm{keV}$\\
\hline\\
     \end{tabular}
     \caption{Key performance requirements for X-IFU}
     \label{xifutable}
  \end{center}
\end{table}
Arrays of  microcalorimeters based on  superconducting transition-edge
sensors (TES) and Frequency Domain Multiplexing (FDM) read out are the
chosen technologies to achieve these ambitious requirements. A TES based
anti-coincidence detector  will be an essential part of the instrument
as well needed to disentangle fake signals produced by high-energy particles.
TES arrays have been used as radiation detectors in a
large variety of wave lengths, from 3-5 mm (as in the CMB experiments \cite{CMB:wiki}) up to  gamma ray \cite{Bennett:2012}.

The X-IFU detector, under development at SRON/VTT and GSFC/NIST, consists of
an   array   of  3840   TES-based   microcalorimeters   read  out   by
Superconducting  QUantum   Interference  Devices  (SQUIDs)   using  96
channels of 40 pixels each  multiplexed in frequency domain in the MHz
bandwidth.  
The  anti-coincidence  detector is part  of the X-IFU
focal plane  assembly and  it is currently under development at
INAF/IAPS-Roma and INFN-Genova \cite{Macculi:2014}.

\section{The TES based detectors}

TES based devices are thermal equilibrium detectors where the incoming
energy is absorbed and converted into equilibrium excitations \cite{Enssbook}. They generally consist of a very sensitive
thermometer coupled to an efficient radiation absorber. The thermometer is made of a
superconducting film operated in the sharp transition between the
normal and the superconducting state where the film resistance varies
between zero and its normal value, typically of the order of tenths to
hundredths of $\mathrm{m}\Omega$. A TES is generally made of thin film
superconducting bi-layer like TiAu, MoAu or MoCu deposited on a SiN thin membrane. The critical
temperature $T_c$ can be accurately chosen by exploiting the proximity
effect between the superconducting (Ti,Mo) and normal (Au,Cu) materials. 

The TES variable resistance is part of a superconducting bias circuit. The
increase of the detector temperature caused by the absorbed radiation
leads to an increase of the TES resistance which is measured as a reduction of the bias current. Voltage biasing of a TES leads to so-called electro-thermal feedback (ETF) \cite{Irwin:1995} which counteracts excursions from the set point. ETF increases the thermal response speed and linearises the
output signal. Biasing can be realised using a dc-voltage source or
using an ac-source (as in the case of FDM)\cite{Kuur:2004}. Typical operation temperatures are around 100 mK.
TES based detectors  can be operated as bolometers to measure power
or as calorimeters to detect energy of single photons. Extremely high
sensitivity has been demonstrated with single pixels: for example, a
dark Noise Equivalent Power of $\sim 1\cdot
10^{-19}\, \mathrm{W}/\sqrt{\mathrm{Hz}}$ with TiAu TES bolometers fabricated at
SRON \cite{Suzuki:2015} and an energy resolution of $1.58\, \mathrm{eV}$ at $6\, \mathrm{keV}$
with MoAu TES micro-calorimeters fabricated at NASA-GSFC
\cite{Smith:ltd14}.
  
\begin{figure}[h]
\centering
\includegraphics[width=0.85\linewidth]{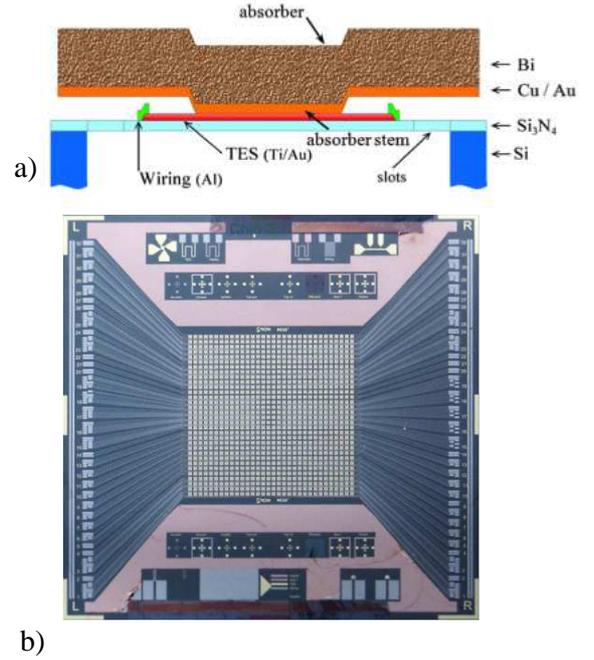}
\caption{{\bf a)} Cross section of a single pixel TES
  microcalorimeter. {\bf b)} a $32\times 32$ TES array fabricated at SRON}
\label{fig:array}
\end{figure}

When operating as a microcalorimeter for soft X-ray spectroscopy, like
in the X-IFU instrument, the TESs are strongly thermally coupled
to a free hanging absorber designed to optimize the quantum efficiency
and the filling factor in a pixels array. The TES-absorber system is
weakly coupled to the thermal bath via thermal conductance $G$. A
schematic diagram of a TES microcalorimeter pixel and an optical
micro-graph of a $32\times 32$ pixels array under development at SRON are shown in figure \ref{fig:array}. 

The energy $E_o$ deposited by a photon in the absorber with heat
capacity $C$ increases the detector temperature of $\Delta
T=E_o/C$. The temperature excursion will then decay exponentially back
to equilibrium with a time constant $\tau=C/G$. The sensitivity of a
TES microcalorimeter  is limited by the thermodynamic energy
fluctuations present in the system. In the small signal limit and in
the case of strong ETF, considering just the thermal-fluctuation and
the TES Johnson noise, the FWHM energy resolution can be approximated
as $\Delta E_{FWHM}=2.335\sqrt{4k_bT^2_c\frac{C}{\alpha}\sqrt{1+2\beta}}$, where
the two parameters $\alpha=T/R\partial R/\partial T$ and $\beta=I/R\partial
R/\partial I$ describe the dependency of the TES resistance on
temperature $T$ and current $I$ \cite{Enssbook}. It becomes clear then that to achieve
high energy resolution one should operate the detector at low
temperature, using low heat capacitance absorber material and a TES
with a steep resistive transition (high $\alpha$) and a low current
dependence (small $\beta$). One should be aware however that a small
heat capacitance and a steep transition can be achieved at the cost of
a reduced detector  dynamic range. An optimisation of the parameters exists as described
in \cite{Ullom:2005}. Typical parameters for TES microcalorimeter are:
$T_C\sim 100\, \mathrm{mK}$, $C(100 \, \mathrm{mK}) \sim 0.5\, \mathrm{pJ/K}$, $\alpha\sim 50-100$ and $\beta\sim 0.2-1$.  

\begin{figure}[h]
\centering
\includegraphics[width=0.9\linewidth]{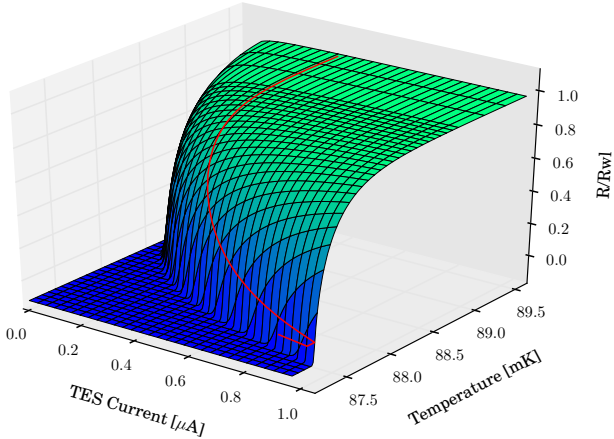}
\caption{Resistive transition $R(T,I)$ of a TES bolometer fabricated
  at SRON  calculated using the resistevely shunted junction model.}
\label{fig:RTI}
\end{figure}

In spite of the fact that TES detectors have been developed for more than
two decades now and are operational in many instruments, the fundamental
physics that describe these devices has only been recently better understood. It has been 
observed that TES devices  behave as superconducting weak-links due to the
longitudinally induced superconductivity  from the superconducting Nb
leads via the proximity effect \cite{Sadleir:2011}. The dependence on the perpendicular magnetic
field coupled into the TES has also been thoroughly  investigated only
in the few past years \cite{Smith:2013}.
By applying the standard resistively shunted model developed for Josephson
junctions and superconducting weak links \cite{Likharev}, it is possible
to calculate the resistive transition R(T,I) of a TES \cite{Kozorezov:2011}. In
figure \ref{fig:RTI} the model of the resistive transition  for a TiAu TES bolometer
developed at SRON \cite{Gottardi:2014} is shown. 
The Josephson current has been directly observed  in TES-bolometers operating in a frequency
domain multiplexer and biased by ac voltage at MHz frequencies \cite{Gottardi:2014}.   
The recent development in understanding the physics of the TES's is
considered as an important step forward in the optimisation of the
performance of a single pixel and a large array of pixels. 

The pixels array for X-IFU, based on MoAu TES's and Au-Bismuth
absorbers is currently under development at NASA-Goddard
\cite{Smith:2015}. SRON is developing the back-up array using TiAu
film technology. 

\section{The Frequency Domain Multiplexing}

The limited cooling power available on a space-based instrument   and
the harness complexity makes the readout of each
individual pixel of the array by its own SQUID amplifier impossible and
SQUID multiplexer schemes such as time domain multiplexing (TDM)
\cite{TDMNIST} and FDM are being developed. TDM is
the most mature technology at the moment \cite{Smith:2015},
however FDM has potentially several important advantages over TDM such as a better use of the
available bandwidth resulting in a larger multiplexing
factor, a smaller thermal load on the cooling chain  and
an easier implementation of the cold electronics on the focal plane assembly. 

SRON-VTT is developing the Frequency Domain Multiplexing and the SQUID
read-out for the X-IFU microcalorimeters array.  FDM requires amplitude modulation
of the TES signal, which can be achieved by biasing the
microcalorimeter with an ac voltage source. Changes of the TES resistance
modulate the amplitude of the ac bias current, following the thermal
signal. In practice the TES signal is moved up to a frequency band
around the carrier signal which is typically in the range between 1
and 5MHz.  Each pixel is separated in the frequency space by high-$Q$
superconducting LC filters \cite{Bruijn:2012}  connected in series with the TES
resistance. The current through each TES is coupled to the SQUID
amplifier input coil through a superconducting summing point. 
The multiplexing factor depends on the available read-out bandwidth and the
detector signal speed. For slow detectors like the low-G bolometers under
development at SRON for the future infrared space mission \cite{Roelfsema:2014}
a multiplexing factor of 160 pixels per SQUID channel has been shown
\cite{Richard:2014sp}. For the faster X-ray microcalorimeters the
current  baseline X-IFU design calls for a multiplexing factor of 40
pixels per SQUID channel which implies a separation of 100 kHz between
each pixel in a read-out bandwidth between 1 to 5 MHz\cite{Hartog:2014sp}.

To improve the limited SQUID dynamic range baseband feedback (BBFB)
\cite{Hartog:2009} is used to cancel the signal at the
sum-point. The signal at the output of the room temperature low noise
amplifier (LNA) is demodulated by digital electronics to extract the
signal  information from the TES, re-modulated and fed back to the SQUID
feedback coil. 
The X-IFU instruments read-out  requires SQUID amplifiers with low
input inductance, low flux noise, high-dynamic range and low power
consumption. We have recently demonstrated nearly quantum-limited
sensitivity with the two-stage SQUIDs amplifiers developed at VTT and
optimised for  the read-out of large array of  infrared bolometers and
X-ray microcalorimeters. We measured an open-input flux noise at $T=25
\mathrm{mK}$ equal to $1.8\cdot  10^{-7}\Phi_0/\sqrt{\mathrm{Hz}}$ and
an input current noise of $1\mathrm{pA}/\sqrt{\mathrm{Hz}}$, which
corresponds to a coupled energy resolution of $\epsilon_{ii} = 15\hbar$ over the whole
interesting frequency range from 2 to 5 MHz \cite{Gottardi:2015}.

\begin{figure}[h]
\centering
\includegraphics[width=0.85\linewidth]{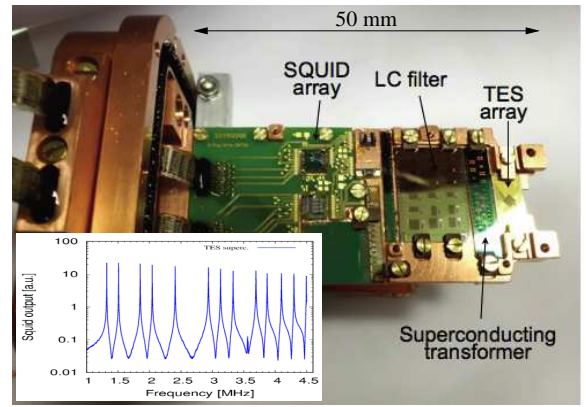}
\caption{Picture of an FDM prototype with VTT two-stage SQUID
  amplifiers, lithographic LC filters and a NASA-Goddard TES
  microcalorimter array. Inset: SQUID output with  13 active resonators.}
\label{fig:FDMprototype}
\end{figure}

An  FDM prototype with 13 active resonators coupled to a NASA-Goddard
TES 8x8 array  is shown in figure \ref{fig:FDMprototype}  

An extended study of the single pixel performance under MHz bias  of the NASA-Goddard TES
microcalorimeters\cite{GSFCTES} is on going at SRON \cite{Akamatsu:2014}. We
have shown an X-ray energy resolution of $2.7\pm 0.2\, \mathrm{eV}$ and
$2.8\pm 0.2\, \mathrm{eV}$ at $5.9\,\mathrm{keV}$ with single pixels read out respectively
at a frequency of 2.3 and $3.7\,\mathrm{MHz}$ \cite{Akamatsu:2015}. 
We are currently optimising a prototype for the first FDM demonstration of GSFC array up to 18 pixels.   
The first demonstration model for the X-IFU instrument with 40 pixels
multiplexing and 4 SQUID channels read-out is scheduled for the coming
year.

\section{The Focal Plane Assembly}

\begin{figure}[h]
\centering
\includegraphics[width=0.75\linewidth]{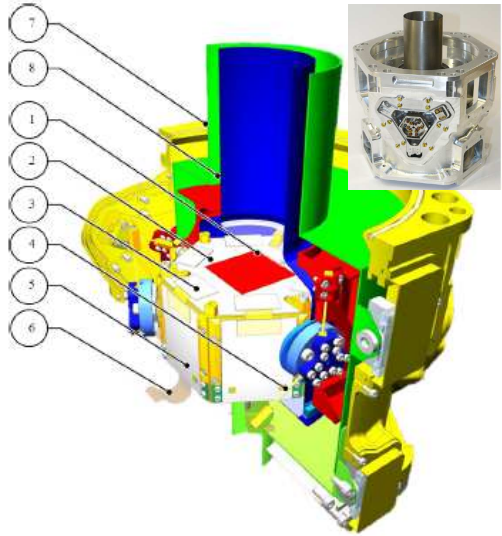}
\caption{FPA cross section with (1) TES array; (2) TES wafer; (3) TES
  wafer to cold electronics unit interconnect; (4) Kevlar thermal
  insulating suspension of 50 mK stage to the 300 mK stage; (5) cold
  electronics unit carrying LC resonators, bias networks and first
  stage SQUIDS; (6) Signal flex cable to higher temperature stage; (7)
  High permeability Cryoperm 10 magnetic shield; (8) Niobium
  superconducting magnetic shield. Inset: Picture of the Nb shield
and its mechanical assembly}
\label{fig:FPA}
\end{figure}

Another challenging development for the X-IFU instrument is the design
and fabrication of the Focal Plane Assembly (FPA) which hosts at a
temperature of $50\mathrm{mK}$  a large-format array of TES's, a large array of lithographic LC
filters, a large amount of SQUID amplifiers and the TES based
anti-coincidence detector. 
The FPA is suspended from a temperature level of $1.7 \mathrm{K}$ to $2 \mathrm{K}$, with an
additional thermal interface at an intermediate temperature of $0.3
\mathrm{K}$. 
Key-technologies are under development such as magnetic shielding, high density
superconducting interconnects and thermal insulating  suspension. A cross-section of the FPA is shown in figure
\ref{fig:FPA} together with a picture of the mechanical assembly of
the Nb shield. A detailed description of the FPA development can be found in \cite{Weers:2014}.
The magnetic shielding design is driven by the following two
requirements: the absolute static normal magnetic field component over
the TES array is less than $10^{-6} \mathrm{T}$ and  the maximum
normal magnetic field noise over the TES array is less than $0.2
\mathrm{nT}/\sqrt{\mathrm{Hz}}$. This  implies having a magnetic shielding
factor as a large as $10^5$ over the whole array size. An on-axis
shielding factor exceeding  $10^{6}$ has been measured in a demonstrator with
a combination of cryoperm and Nb shields. \cite{Weers:2014}.

We are currently developing  a novel superconductive technology for
the interconnection between the TES array and the cold electronics using
transformed based coil coupling, an unique feature of the frequency
domain read-out. If proven successful this technique will allow
high-density connections  and easy
replacement of the pixels array during the development and engineering phase of
this very complicated space instrument. In the current design each
pixel is fanned-out to a superconducting  coil at the edge of the TES
wafer. The mating interconnect pieces contain the secondary coils
which form a transformer once assembled. They are fabricated by
spinning and patterning of polyimide on a Si wafer after that  Nb
tracks  are deposited using standard lithographic processing. In a
final DRIE step the interconnect are etched from the rest of the
Si wafer leaving a flexible free standing membrane
\cite{Weers:2014}. 

The mechanical support of the detector stage  and the 300 mK level has
to fulfil stringent  requirements on position stability of the
detector array over multiple cool downs, compactness in terms of
space, modularity in assembly  and sufficient thermal isolation between
the stages. A kinematic suspension design using Kevlar cords have
been chosen and is currently under development and test \cite{Weers:2014}.

\section{Conclusion}
{\it Athena} has been selected by ESA as its second large-class mission. The future European X-ray observatory will study the hot and energetic Universe with its launch foreseen in 2028. 
Microcalorimeters based on superconducting  Transition-edge sensor
(TES)  are the chosen technology for the detectors array of the X-IFU instrument on board of Athena
SRON/VTT, GSFC/NIST and INAF(Italy) are developing the first
demonstration model for the X-IFU instrument with 40 pixels
multiplexing and 4 SQUID channels read-out and the anti-coincidence
detector. The extended activities include the
fabrication and the testing of the large array of  high energy resolution X-ray TES
microcalorimeter, the cold electronics for the frequency domain
multiplexing, the TES based anti-coincidence detector and the key-technologies for the focal plane assembly.

\section*{Acknowledgements}

H.A. was supported by a Grant-in-Aid for Japan Society for the Promotion of Science (JSPS) Fellows (22-606)



\begin{thebibliography}{26}
\expandafter\ifx\csname url\endcsname\relax
  \def\url#1{\texttt{#1}}\fi
\expandafter\ifx\csname urlprefix\endcsname\relax\def\urlprefix{URL }\fi
\bibitem{Barcons:2015} X.~Barcons, et~al.,J. Phys:Conf.Ser 610 (2015) 012008 
\bibitem{Ravera:2014} L.~Ravera, Proc. SPIE 9144, (2014) 91442L.
\bibitem{CMB:wiki} {\it List of cosmic microwave background experiments}, Wikipedia, The Free Encyclopedia, (2015).

\bibitem{Bennett:2012} D.~Bennett, et-al., Rev. Sci. Instr. 83 9
  (2012) 093113. 
\bibitem{Macculi:2014} C.~Macculi, et-al, Proc. SPIE 9144,  (2014) 91445S.
\bibitem{Enssbook} Ch.~Enss, {Cryogenic Particle Detection}
  (Springer-Verlag,2005)
\bibitem{Irwin:1995} K.~Irwin, Appl. Phys. Lett. 66 (1995) 1998.
\bibitem{Kuur:2004} J.~van der Kuur, et~al., Nucl. Instr. Methods Phys. Res. A 520 (2004) 551.
\bibitem{Suzuki:2015} T.~Suzuki to be published
\bibitem{Smith:ltd14} S.~J. Smith, et~al., J. Low Temp. Phys. 167 (2012)
  168-175. 
  
\bibitem{Ullom:2005} J.~N. Ullom, et~al, Appl. Phys. Lett., 87 (2005) 194103. 
\bibitem{Sadleir:2011} J.~Sadleir, et~al, Phys. Rev. B 84 (2011)
  184502.
\bibitem{Smith:2013} S.~J. Smith, et~al., J. Appl. Phys. 114 (2013) 074513.
\bibitem{Likharev} K.~Likharev, Rev. Mod. Phys. 51, (1979) 101. 
\bibitem{Kozorezov:2011} A.~Kozorezov, et~al., Appl. Phys. Lett. 99 (2011) 063503. 
\bibitem{Gottardi:2014} L.~Gottardi, et~al., Appl. Phys. Lett. 105
  (2014) 162605.
\bibitem{Smith:2015} S.~J. Smith, et~al., IEEE Trans. Appl. Superc. (2015).
\bibitem{TDMNIST} W.~B. Doriese, et~al, J. Low Temp. Phys 167 (2012) 595-601.
\bibitem{Bruijn:2012} M.~Bruijn, et~al., J. of Low Temp. Phys. 167
  (2012) 695700
\bibitem{Roelfsema:2014} P.~Roelfsema, Proc. SPIE 9143 (2014) 91431K.
\bibitem{Richard:2014sp} R.~A. Hijmering, Proc. SPIE. 9153 (2014) 91531E.
\bibitem{Hartog:2014sp}  R~.H. den Hartog, et al., Proc. SPIE 9144 (2014) 91445Q . 
\bibitem{Hartog:2009} R.~den Hartog, et~al.,AIP Conf. Proc. 1185
  (2009) 261 
\bibitem{Gottardi:2015} L.~ Gottardi, et~al., IEEE Trans. Appl. Superc. (2015).
\bibitem{GSFCTES} N.~Iyomoto, et~al., Appl. Phys. Lett. 92 (2008) 13508 and S.~R. Bandler, et~al., J. Low Temp. Phys. 151 (2008) 400405
\bibitem{Akamatsu:2014} H.~Akamatsu, et~al., J. of Low
  Temp. Phys.,(2014) 1–6 
\bibitem{Akamatsu:2015} H.~Akamatsu, et~al., to be published in J. of Low Temp Phys.

\bibitem{Weers:2014} H.~J. van Weers, et~al., Proc. SPIE 9144  (2014)
  91445R and A.~Bergen, et~al., submitted to RSI (2015) 




\end{thebibliography}
\end{document}